\documentclass{ws-procs9x6}
\begin{document}

\title{ The measurement, fitting and interpretation of the
radial distributions of Heavy-Light
mesons calculated on a lattice  with dynamical fermions.}
\author{UKQCD Collaboration}

\author{A.M. Green, J. Koponen, P. Pennanen}

\address{Department of Physical Sciences and Helsinki Institute of
Physics\\
P.O. Box 64, FIN--00014 University of Helsinki, Finland\\
E-mail: anthony.green@helsinki.fi}
\author{C. Michael}

\address{Department of Mathematical Sciences, University of Liverpool,
 L69 3BX, UK\\
E-mail: cmi@liv.ac.uk}
\maketitle
\begin{figure}[ht]
\centering
\includegraphics*[width=0.7\textwidth]{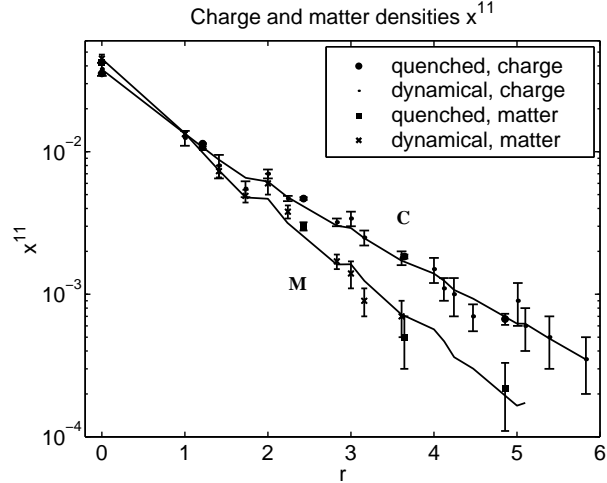}
\caption{The ground state charge (C) and matter (M) densities
[$x^{11}(r)$] as a
function of $r/a$ from Ref.~\protect\cite{G+K+P+M2}.
The lines shows a fit to these densities with a sum of
two lattice exponential functions.
The scaled quenched results of
Ref.~\protect\cite{G+K+P+M} are also shown by filled circles and
squares.}
\label{x11vs2LY}
\end{figure}

\thispagestyle{empty}
\abstracts{In our earlier work, the charge and matter radial
distributions of
heavy-light mesons were measured on a $16^3\times 24$ lattice with
a lattice spacing of $a \approx$ 0.17 fm and a light quark mass about
that of the strange quark.\\
Several major improvements have now been made:\\
1) Dynamical fermions are  used  with $a \approx$ 0.14 fm;\\
2) More gauge configurations are included (78 vs 20);\\
3) Off-axis, in addition to on-axis, insertions are made;\\
4) The data analysis is much more complete. In particular, distributions
involving excited states are extracted.}
\noindent Using the lattice parameters given in the abstract, the charge and matter
radial distributions of a heavy-light meson are as shown in
Fig.\ref{x11vs2LY} in terms of
$x^{ij}(r)=\langle \psi^i|\Theta(r)|\psi^j\rangle$, where the operator
$\Theta(r)$ measures either the charge or the matter distribution of
the light quark at distance $r$ from the heavy quark.
Several points are of interest:
\begin{itemize}
\item[1)] At small values  of $r$ the two densities are comparable i.e.
$x^{11}_C(0) \approx x^{11}_M(0)$.
\item[2)] As $r$ increases, the matter density decreases faster
than the charge density.
\item[3)] The densities calculated with the quenched approximation in
Ref.~\cite{G+K+P+M} are the same --- within error bars --- as those
for the full dynamical quark calculation of Ref.~\cite{G+K+P+M2}.
\item[4)] The densities do not have a smooth variation with $r$.
This is expected, since the lattice discretization leads to a
breaking of rotational invariance and the appearance of kinks in
the densities.
\end{itemize}

\begin{figure}[ht]
\centering
\includegraphics*[width=0.99\textwidth]{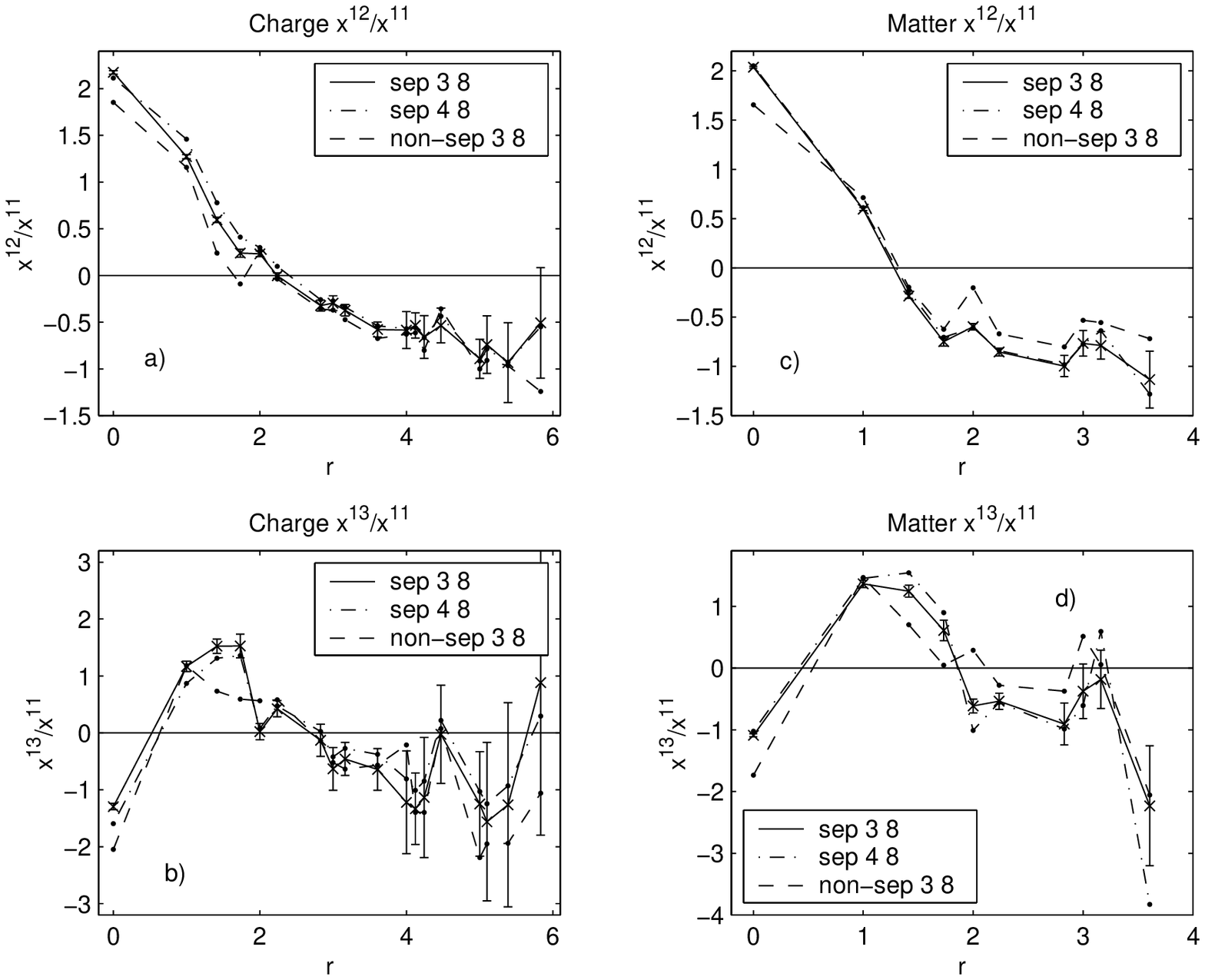}
\caption{a) and b): The ratios $x^{12}/x^{11}$ and  $x^{13}/x^{11}$
for the charge distribution.
\mbox{c) and d): The ratios $x^{12}/x^{11}$ and  $x^{13}/x^{11}$
for the matter distribution.}}
\label{x12x11}
\end{figure}

 Figure~\ref{x12x11} shows the results for the
first and second excited states.
 The presence of  nodes is very clear with the  first excited state
having a single node at $\approx  0.3$ fm
in the charge case and $\approx  0.2$ fm for the matter.
The second excited state seems to have  two nodes with one being at
$\approx  0.1$ fm and a second at $\approx  0.4$ fm for the charge and 0.3 fm
for the matter. It will be a challenge for phenomenological models to
explain this data.

\begin{table}[b!]
\centering
\tbl{Masses of exchanged particles in GeV.}
{\footnotesize
\begin{tabular}{|c|cc|}
\hline
&$M^{V}$&$M^{S}$ \\ \hline
Lattice Yukawa fit to $Q\bar{q}$ data&0.78(3)&1.53(8) \\
Lattice Exponential fit to $Q\bar{q}$ data&1.00(6)&1.50(7) \\
Direct calculation as $q\bar{q}$ states~\protect\cite{Allton}&1.10(13)&1.65(11)\\
\hline
\end{tabular}}
\end{table}

These distributions can be fitted well with latticized Yukawa or
exponential forms. If, in each case, two such forms are used, then
the longer ranged part can be expressed as the exchange of a meson of
mass $M^{V}$ ($M^{S}$) for the charge (matter) distribution.
These masses can be compared with the values obtained directly as
$q\bar{q}$ states ~\cite{Allton}.

\end{document}